\documentclass[10pt]{article}

\usepackage{cite}

\usepackage{graphicx}        
\usepackage{amssymb}
\usepackage{amsmath}
\usepackage{pdflscape}
\usepackage{tikz}
\usepackage[scriptsize]{subfigure}
\usepackage{amsmath}
\usepackage{soul}

\usepackage{float}

\usepackage[T1]{fontenc}
\usepackage{kpfonts,baskervald}

\usepackage[utf8]{inputenc}
\usepackage{authblk}

\usepackage{placeins}

\usepackage{caption} 
\usepackage{dblfloatfix} 

\tikzset{cross/.style={cross out, draw=black, minimum size=2*(#1-\pgflinewidth), inner sep=0pt, outer sep=0pt},
cross/.default={2.5pt}}
\usepackage[margin=1in]{geometry}
\usepackage{parskip}

\DeclareGraphicsExtensions{.pdf,.png,.jpg,.eps} 
\usepackage{epstopdf}
\usepackage[acronym,shortcuts]{glossaries}

\makeglossaries

\newacronym{GFVSC}{\textcolor{black}{GF-VSC}}{Grid-forming voltage source converter}

\begin{document}
%
\title{Coordinated control in multi-terminal VSC-HVDC systems to improve transient stability: Impact on electromechanical-oscillation damping}
\date{ }

\author[1]{Javier Renedo} \author[2]{Luis Rouco} \author[2]{Aurelio Garcia-Cerrada} \author[2]{Lukas Sigrist}

\affil[1]{ETSI ICAI, Universidad Pontificia Comillas,  Madrid, Spain}
\affil[2]{Instituto de Investigaci\'on Tecnol\'ogica (IIT), ETSI ICAI, Universidad Pontificia Comillas,  Madrid, Spain}

\maketitle

\section*{Abstract}\label{sec:abstract}
Multi-terminal high-voltage Direct Current technology based on Voltage-Source Converter stations (VSC-MTDC) is expected to be one of the most important contributors to the future of electric power systems. In fact, among other features, it has already been shown how this technology can contribute to improve transient stability in power systems by the use of supplementary controllers. Along this line, this paper will investigate in detail how these supplementary controllers may affect electromechanical oscillations, by means of small-signal stability analysis. The paper analyses two control strategies based on the modulation of active-power injections (P-WAF) and reactive-power injections (Q-WAF) in the VSC stations. Both control strategies use global signals of the frequencies of the VSC-MTDC system and they presented significant improvements on transient stability. The paper will provide guidelines for the design of these type of controllers to improve both, large- and small-disturbance angle stability. Small-signal stability techniques (in Matlab) will be used to assess electromechanical-oscillation damping, while non-linear time domain simulation (in PSS/E) will be used to confirm the results. Results will be illustrated in Nordic32A test system with an embedded VSC-MTDC system. The paper analyses the impact of the controller gains and communication latency on electromechanical-oscillation damping. The main conclusion of the paper is that transient-stability-tailored supplementary controllers in VSC-MTDC systems can be tuned to damp inter-area oscillations too, maintaining their effectiveness for transient-stability improvement.

\noindent \textbf{Index terms:} VSC-HVDC, multi-terminal, transient stability, electromechanical oscillations, power oscillation damping (POD)

\noindent Internal reference of this article: IIT-22-105WP

\noindent Contact information: javier.renedo@ieee.org, luis.rouco@iit.comillas.edu, aurelio@iit.comillas.edu, lukas.sigrist@iit.comillas.edu.



\newpage

\section{Introduction}\label{sec.Intro}

\noindent Multi-terminal high voltage direct current systems based on voltage source converters (VSC-HVDC) is a key technology for bulk power transmission and for the integration of renewable resources into power systems. Although it is clear that the main aplication of VSC-HVDC systems is power transmission, they can also help to improve the operation of power systems by means of supplementary controllers. Previous publications have proposed supplementary controllers in multi-terminal VSC-HVDC systems (VSC-MTDC, for short) to improve angle stability against small disturbances (electromechanical oscillations or power-oscillation damping, POD) and against large disturbances (transient stability).  

The work in~\cite{Harnefors2014} proposed supplementary controllers in VSC-MTDC systems to damp electromechanical oscillations (POD controllers). In the control strategy, the VSCs modulate their active-power (P) injections using the information of the speed of some generators of the system, using a Wide Area Measurement System~(WAMS). The work in~\cite{Eriksson2014b} proposed POD controllers in DC-voltage-droop-controlled VSC-MTDC systems. In this control strategy, one converter of the VSC-MTDC changes its DC-voltage set point proportionally to the locally measured frequency deviation. The paper uses the concept of DC-voltage loop shaping.  The work in~\cite{Preece2014} proposed POD controllers for P injections of the VSC stations, using the active-power flows through a set of AC lines as input signal~(using a WAMS). The work in~\cite{iit_vsc_mtdc_pod2021} proposed POD controllers in VSC-MTDC systems by modulating P and reactive-power~(Q) injections of the VSCs and using global measurements of the frequencies at the connection points of the VSC stations of the MTDC system. Reference~\cite{Marinescu2021} proposed POD controllers in a VSC-HVDC link embedded in an AC grid based on Linear Matrix Inequality~(LMI) optimisation and modulating P/Q injections at the converter stations.  

The improvement of transient stability of a power system using a VSC-MTDC system has already been addressed in the literature. The work in~\cite{Eriksson2014a} proposed a control strategy in VSC-MTDC systems for transient-stability improvement, where VSC stations control their P injections based on a bang-bang controller and using the speed deviations of the generators with respect to the speed of the centre of inertia (COI) as input signals. Reference~\cite{Tang2016} proposed controlling P injections of the VSC stations of a MTDC system based on sliding-mode strategy and also using global mesurements. The work in~\cite{javierr2016} proposed a control strategy where VSC stations controlled their P injections, using global measurements of the frequencies of the MTDC system. The work in~\cite{iitcontrolQ2017} used the same input signal, but to control the Q injections of the VSC stations. The work in~\cite{JuanCarlos_transient_stab2021} proposed an active-power control strategy in VSC-MTDC systems using global measurements of the angles and frequencies at the connection point of the VSC stations.  

In general, small- and large-signal angle stability are related and the latter often improve the former too, and vice-versa. However, this is not a always the case~\cite{Kundur1989}. For example, the study presented in~\cite{Kundur1989} reported a case in which an increase in the gain of the PSS of a synchronous generator results in an increase of the damping ratios of electromechanical modes, but in a deterioration of transient stability. Hence, the following questions  remain open:
\begin{itemize}
	\item What is the impact of transient-stability-tailored controllers in VSC-MTDC systems on electromechanical-oscillation damping?
	\item Could transient-stability-tailored controllers in VSC-MTDC systems help to damp electromechanical oscillations too and play the role of POD controllers too?
\end{itemize}

Along this line, this paper studies the impact of transient-stability-tailored supplementary controllers in VSC-MTDC systems on electromechanical-oscillation damping by means of small-signal stability analysis. The paper analyses two control strategies based on the modulation of active-power injections (P-WAF)~\cite{javierr2016} and reactive-power injections (Q-WAF)~\cite{iitcontrolQ2017}  in the VSC stations of the MTDC system. Both control strategies use global signals of the frequencies of the VSC-MTDC system and they presented significant improvements on transient stability. Results will suggest guidelines for the design of these type of controllers to improve both, large- and small-disturbance angle stability. Small-signal stability analysis techniques (in Matlab) will be used to assess electromechanical-oscillation damping, while non-linear time domain simulation (in PSS/E) will be used to confirm the results. Contributions will be illustrated in the Nordic32A test system with an embedded VSC-MTDC system. The paper analyses the impact of the controller gains and communication latency on electromechanical-oscillation damping. 

Preliminary results were presented by the authors in~\cite{jrenedoSSA2019}, where small-signal stability analysis of strategy P-WAF was carried out in a small test system. This paper extends the results an analyses not only the modulation of P injections (P-WAF), but also the modulation of Q injections (Q-WAF) and simultaneous modulation of P and Q injections (PQ-WAF). Furthermore, this paper presents the results in a larger test system (Nordic32A test system) and it analyses the impact of communication latencies on the performance of the control strategies, which was not analysed in~\cite{jrenedoSSA2019}.

\section{Multi-terminal VSC-HVDC systems}\label{sec.vsc_mtdc_systems}

\begin{figure}[!htbp]
\begin{center}
\includegraphics[width=1\columnwidth]{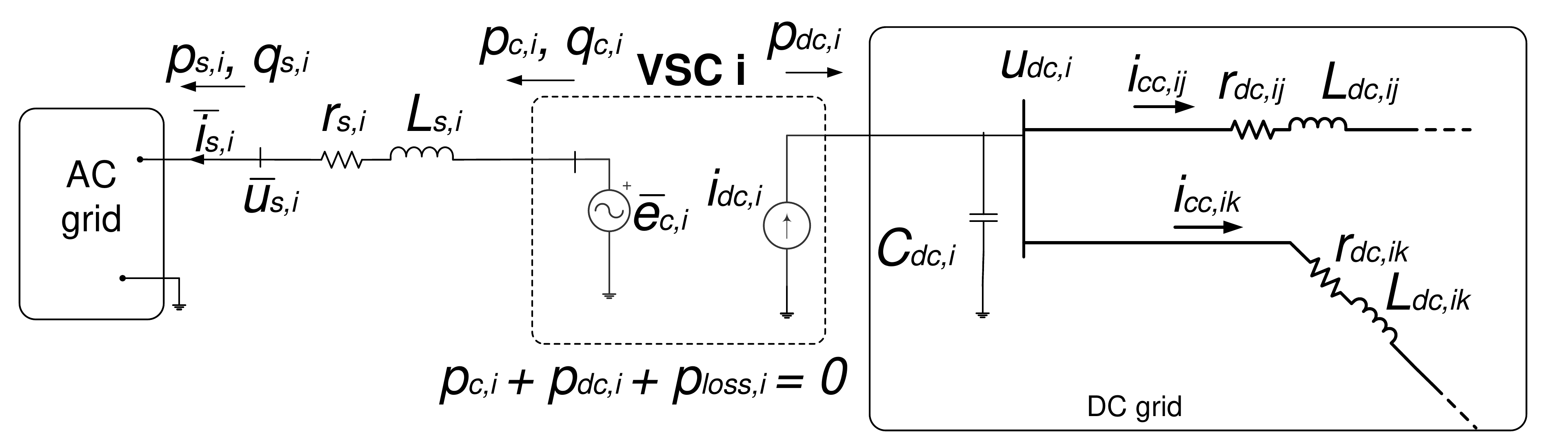}
\caption{VSC and the HVDC grid based on the principles of~\protect\cite{Cole2010}.}
\label{fig:vsc_dcgrid_model}
\end{center}
\end{figure}

\noindent A VSC-MTDC system consists of $n$ VSC stations connected to the same HVDC grid. Fig.~\ref{fig:vsc_dcgrid_model} shows the dynamic model of a VSC connected to an HVAC grid and to an HVDC grid, following the guidelines of~\cite{Cole2010, Beerten2014,Liu2014} for electromechanical-type models of this type of system (type-6 models, according to the classification of \cite{Saad2013}). 

A VSC is typically controlled by using vector control. In this paper, the outer controllers are represented in detail, whereas the closed-loop system of the inner current loops are represented as first order systems. This is reasonable for electromechanical-type studies where the dynamics of interest are slow in comparison with the dynamics of the inner current loops. Converter limits are implemented in the model: P/Q limits, current limit and maximum modulation index. Converter losses are calculated as proposed in~\cite{Daelemans2009}:
\begin{equation}\label{eq:VSC_ploss}
p_{loss,i} = a_i + b_i \cdot i_{s,i} + c_i \cdot i_{s,i}^2
\end{equation}
where $i_{s,i}$ is the magnitude of the current injection of the VSC in pu (RMS).

Each VSC station is seen by the HVDC grid as a current injection ($i_{dc,i}$). The HVDC grid contain buses and lines. DC buses have an equivalent capacitance~($C_{dc,i}$) connected that aggregates the equivalent DC-link capacitor of the VSC and the contribution of the shunt capacitances of lines connected to the bus. DC lines are modelled as series resistances~($r_{dc,ij}$) and series inductances~($L_{dc,ij}$). 

In this paper, the initial operating point is calculated with the sequential AC/DC power flow algorithm proposed in~\cite{Beerten2012}. All the details of the dynamic model used in this work can be found in~\cite{jrenedoPSSE2017}. For small-signal stability analysis, the system must be linearised. All the details of the linearised model used in this work can be found in~\cite{jrenedoSSA2019}.

\FloatBarrier
\newpage
\section{Control strategies}\label{sec.vsc_mtdc_control_strat}
\noindent The control strategies to be investigated were proposed in~\cite{javierr2016,iitcontrolQ2017} and this paper will analyse their impact on electromechanical-oscillation damping. In these control strategies, each VSC station of the MTDC system compares its own frequency measured at the connection point with a frequency set point calculated as the weighted average of the frequencies measured at the AC side of the VSC stations~(weighted-averaged frequency, WAF): 
\begin{equation}\label{eq.weightedavinitial}
 \omega^* = \bar{\omega} = \sum_{k=1}^n \alpha_k \omega_k \mbox{(pu) \space with \space}  \alpha_k \in [0,1] \mbox{ \space and } \sum_{k=1}^n \alpha_k = 1. 
\end{equation}
where $\omega_k$ is the frequency measured at the connection point of VSC$_{k}$.

The frequency error signal is used by each VSC to modulate its P Injection (strategy P-WAF) and/or its Q injection (strategy Q-WAF). 

\subsection{Strategy P-WAF}\label{sec.control_strategies_PWAF}
\begin{figure}[!htbp]
\begin{center}
\includegraphics[width=0.8\columnwidth]{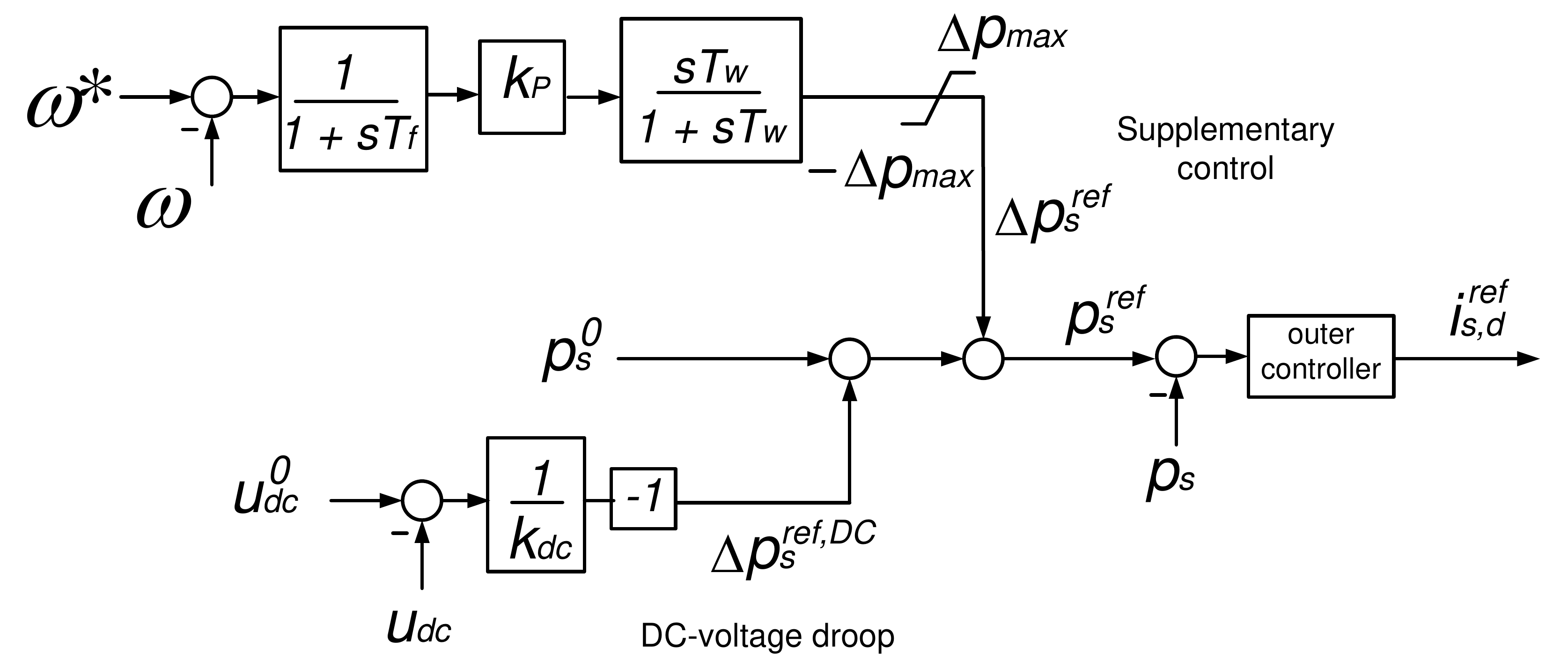}
\caption{Control for the active-power injection of a VSC in strategy P-WAF. Frequency in pu.}
\label{fig:P_supp_control}
\end{center}
\end{figure}

\noindent Fig.~\ref{fig:P_supp_control} shows the block diagram of control strategy P-WAF~\cite{javierr2016}. A supplementary set point ($\Delta p_{s}^{ref}$) is added to the active-power set point of each VSC. The controller has a proportional gain~($k_{P}$), a low-pass filter for noise filtering (with time constant~$T_{f}$), a wash-out filter (with time constant~$T_{W}$)  and a saturation parameter~($\Delta p_{max}$). The frequency set point is calculated as the WAF in~(\ref{eq.weightedavinitial}). In Fig.~\ref{fig:P_supp_control}, the VSC has the DC-voltage droop control~($\Delta p_{s}^{ref,DC}$) implemented together with the control strategy, but this option is not mandatory. 

The behaviour of strategy P-WAF is as follows. During the transient produced by a disturbance, the frequencies seen the VSCs will change. If a VSC has the frequency above (below) the WAF, it will decrease (increase) its P injection, aiming to pull together the speeds of the generators of the system.

The work in~\cite{javierr2016} proposed the design of the controller gains proportionally to the weighting factors used to calculate the WAF:
\begin{equation}\label{WAF_alpha}
	 \frac{k_{P,k}}{k_{P,T}} =  \alpha_k, \mbox{ \space with \space }   k_{P,T}=\sum_{j=1}^n  k_{P,j}.
\end{equation} 
This design ensures $ \sum_{j=1}^n\Delta p^{ref}_{s,j} = 0$ and avoids the interaction of the supplementary controller with the DC-voltage droop control. A reasonable design of the weighting factors ($\alpha_k$) is to select them proportionally to the rating of each VSC$_k$ of the MTDC system.

\FloatBarrier

\subsection{Strategy Q-WAF}\label{sec.control_strategies_QWAF}
\begin{figure}[!htbp]
\begin{center}
\includegraphics[width=0.7\columnwidth]{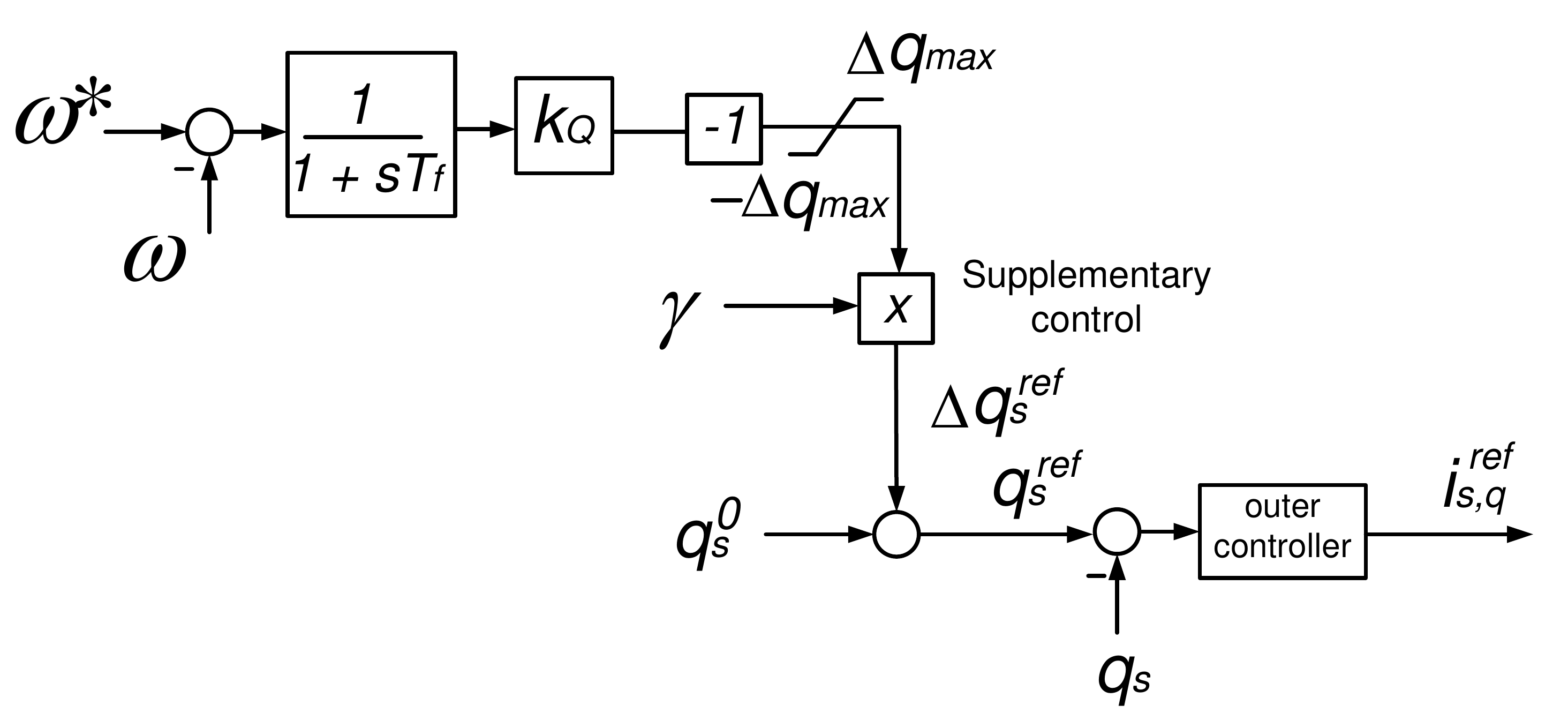}
\caption{Control for the reactive-power injection of a VSC in strategy Q-WAF. Frequency in pu.}
\label{fig:Q_supp_control}
\end{center}
\end{figure}

\noindent Fig.~\ref{fig:Q_supp_control} shows the block diagram of control strategy Q-WAF~\cite{iitcontrolQ2017}. A supplementary set point ($\Delta q_{s}^{ref}$) is added to the reactive-power set point of each VSC. The controller has a proportional gain~($k_{Q}$), a low-pass filter for noise filtering (with time constant~$T_{f}$), a wash-out filter (with time constant~$T_{W}$), a saturation parameter~($\Delta q_{max}$) and a binary variable for the activation of the controller ($\delta$). The controller is activated  if the AC voltage at the connection point is above a certain threshold $V_{TH}$ ($\delta=1$ if $u_s\ge V_{TH}$). This condition is used to produce the control actions only after the fault clearing (preventing the controller acting during the short circuits), which improves the performance of the controller. The frequency set point is calculated as the WAF~(\ref{eq.weightedavinitial}).

The behaviour of strategy Q-WAF is as follows. If a VSC has the frequency above (below) the WAF, it will increase (decrease) its Q injection, aiming to pull together the speeds of the generators of the system. Notice that the block diagram of Fig.~\ref{fig:Q_supp_control} contains a negative sign.

The work in~\cite{iitcontrolQ2017} used the following design:
\begin{equation}\label{QWAF_alpha}
	\frac{k_{Q,k}}{k_{Q,T}} = \alpha_k \mbox{, \space}  k_{Q,T}=\sum_{j=1}^n k_{Q,j}.
\end{equation}
which ensures $\sum_{j=1}^n \Delta q^{ref}_{s,j} = 0$, analogously to strategy P-WAF. However, this condition is not mandatory for control of Q injections.

\FloatBarrier

\newpage

\section{Results}\label{sec.results2}

\begin{figure}[!htbp]
\begin{center}
\includegraphics[width=0.80\columnwidth]{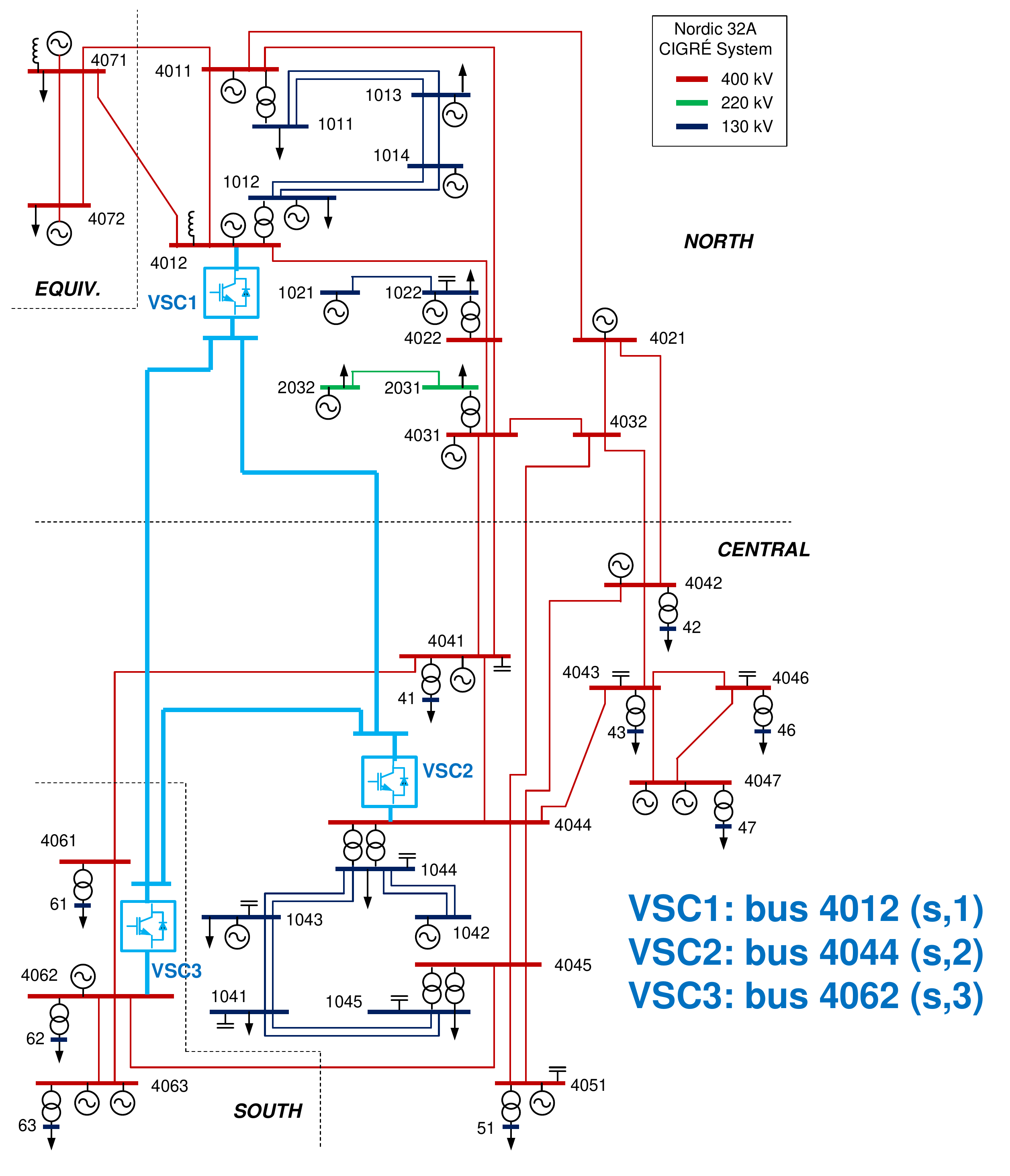}
\caption{Nordic32A system with an embedded VSC-MTDC.}
\label{fig:Nordic32A_mtdc}
\end{center}
\end{figure}

\noindent The case study considered consists of the Cigr\'e Nordic32A test system~\cite{CIGREB_Nordic32A_1995} with an embedded 3-terminal VSC-HVDC system, as shown in Fig.~\ref{fig:Nordic32A_mtdc}. Each VSC has a rating of 1000~MVA. A critical scenario with poorly damped inter-area oscillations is considered. The modifications made to stress the system are provided in~\ref{sec.app_test_system}. The data of the VSC-MTDC system can also be found in~\ref{sec.app_test_system}.

Each VSC of the MTDC system is controlled with DC-voltage droop control and constant reactive power injection. The initial steady-state operating point of the VSC-MTDC system has been obtained with an AC/DC power flow \cite{Beerten2012,jrenedoPSSE2017}, with the following specified variables:
\begin{itemize}
	\item VSC1: $P_{s,1}^{0} = -350$ MW and $Q_{s,1}^{0} = 0$ MVAr.
	\item VSC2: $P_{s,2}^{0} = 500$ MW and $Q_{s,2}^{0} = 150$ MVAr.
	\item VSC3: $u_{dc,3}^{0} = 1$ pu and $Q_{s,3}^{0} = 100$ MVAr  (VSC3 is the DC-slack converter for power-flow calculation).
\end{itemize}

The dynamic model of the system in Fig.~\ref{fig:Nordic32A_mtdc} has been linearised around the steady-state operation point and the linearised model has been implemented in Matlab-based Small Signal Stability Tool (SSST)~\cite{SSST_manual}, as described in~\cite{jrenedoSSA2019}.

Electromechanical oscillations have been identified using eigenvalue techniques and participation-factors~\cite{IPA1982a,Pagola1989}. The system has two inter-area modes with low damping ratios, as depicted in Table~\ref{tab:Nordic32A_modes_electromechanical}. 

\begin{table}[!htbp]
\begin{center}
\caption{Inter-area modes. N: North, S: South, C: centre}
\label{tab:Nordic32A_modes_electromechanical}
\scalebox{0.90}{
\begin{tabular}{|c|ccc|c|c|}
\hline
Mode &Eigenvalue & $\zeta$ & Freq.  & Machines  & Oscillation \\ 
 & (rad/s) & (\%) & (Hz) &   with greatest &  \\ 
 &  &  &  &   participation &  \\ \hline 
A & $-0.1044 \pm j 3.2333$ & 3.23 & 0.51& G4072, G4063 & N - S \& C\\
B & $-0.3186 \pm j 5.2160$ & 6.10 & 0.83 &  G4063, G4072,   & N \& S - C \\ 
\null & \null & \null & \null & G1042 & \null \\
\hline
\end{tabular}
}
\end{center}
\end{table}

\begin{figure}[!htbp]
\begin{center}
\includegraphics[width=0.6\columnwidth]{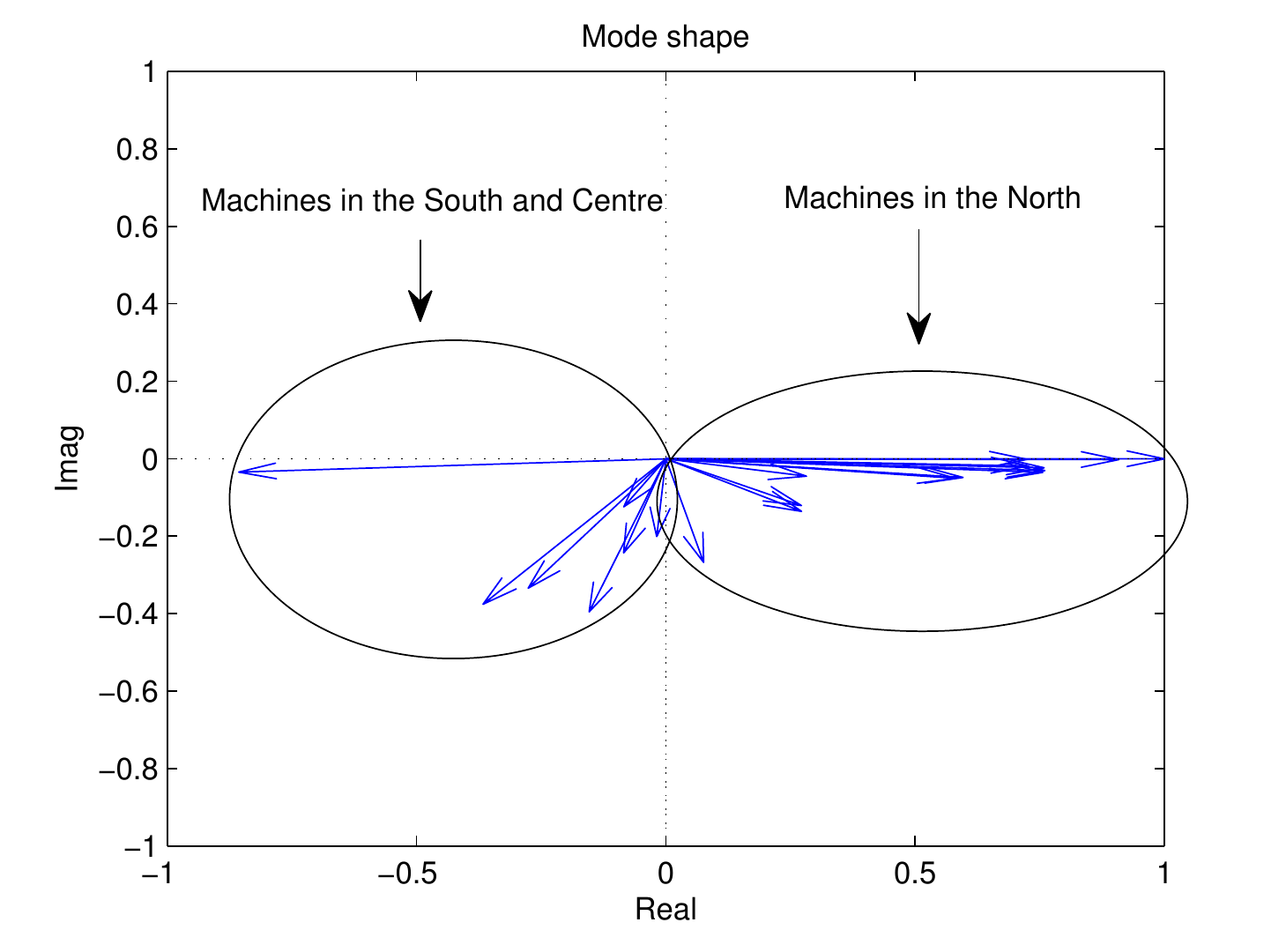}
\caption{Shapes of inter-area mode A.}
\label{fig:Nordic32A_modeA_shape}

\includegraphics[width=0.6\columnwidth]{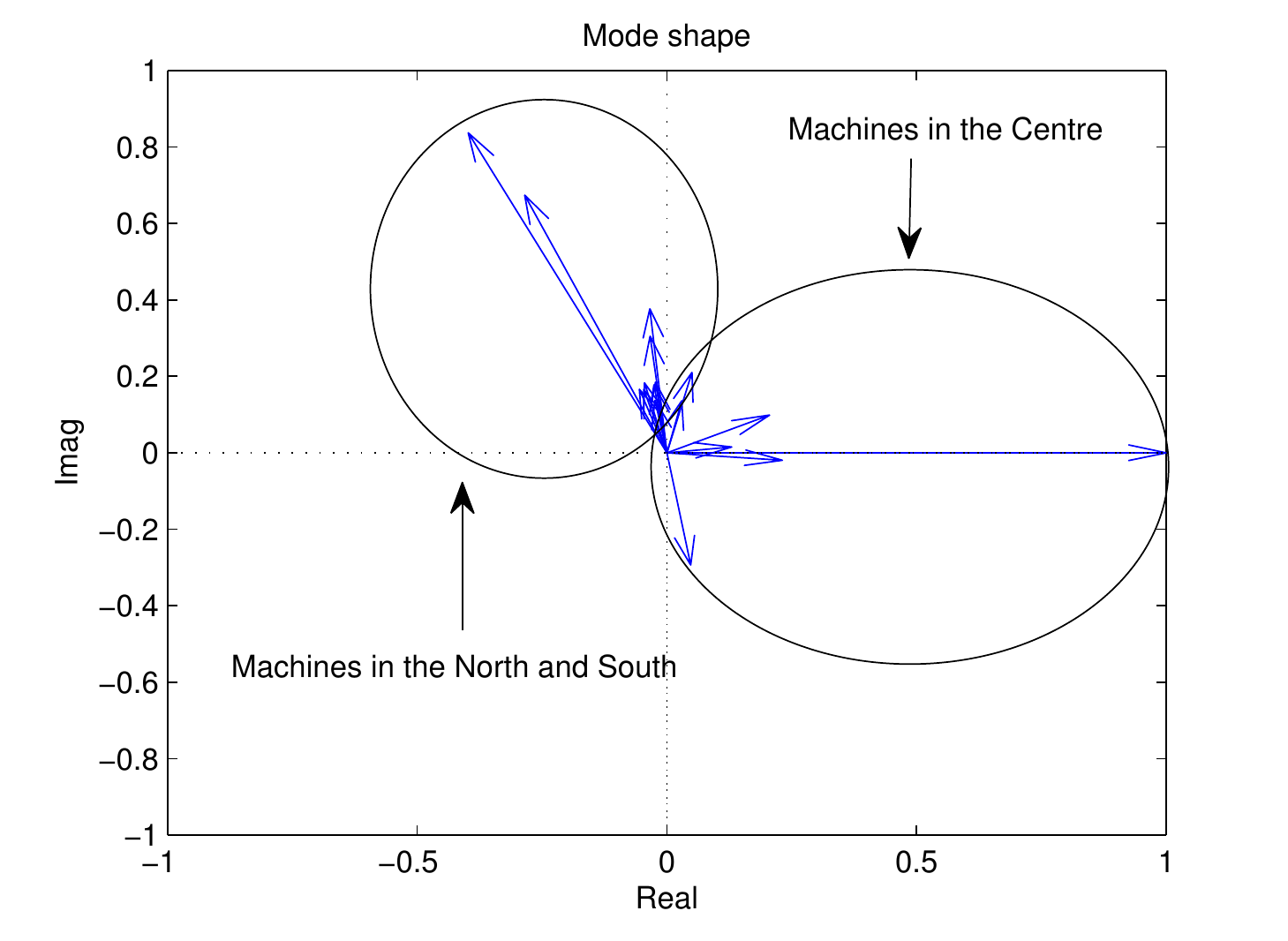}
\caption{Shapes of inter-area mode B.}
\label{fig:Nordic32A_modeB_shape}
\end{center}
\end{figure}

Figs.~\ref{fig:Nordic32A_modeA_shape} and~\ref{fig:Nordic32A_modeB_shape} show the shapes of inter-area modes A and B  (right eigenvectors associated to the speeds of the synchronous machines), respectively. In inter-area mode A, synchronous machines in the North oscillate against synchronous machines in the South and in the Centre. This can be seen in the phase of the mode shapes: mode shapes of the speeds of the generators in the North have opposite phase than mode shapes of the speeds of the generators in the South. In inter-area mode B, synchronous machines in the North and South oscillate against machines in the Centre. 


\FloatBarrier

\subsection{Small-signal stability analysis}\label{sec.results2_ssa}

\noindent The impact of the controller gains of strategies P-WAF (P injections) and Q-WAF (Qinjections) on inter-area modes is analysed. Gains at all VSC stations are changed (satisfying~(\ref{WAF_alpha}) and (\ref{QWAF_alpha})) and eigenvalues and their damping ratio are obtained. Since gains  in the range $k_{P,i}, k_{Q,i}=[100,300]$~pu (pu's are referred to thc converter rating) proved to be effective to improve transient stability~\cite{javierr2016,iitcontrolQ2017}, gains are analysed in the range $k_{P,i}, k_{Q,i}=[0,500]$~pu. The rest of parameters of the control strategies are provided in~\ref{sec.control_strat_param}.

\begin{figure}[!htbp]
\begin{center}
\includegraphics[width=0.6\columnwidth]{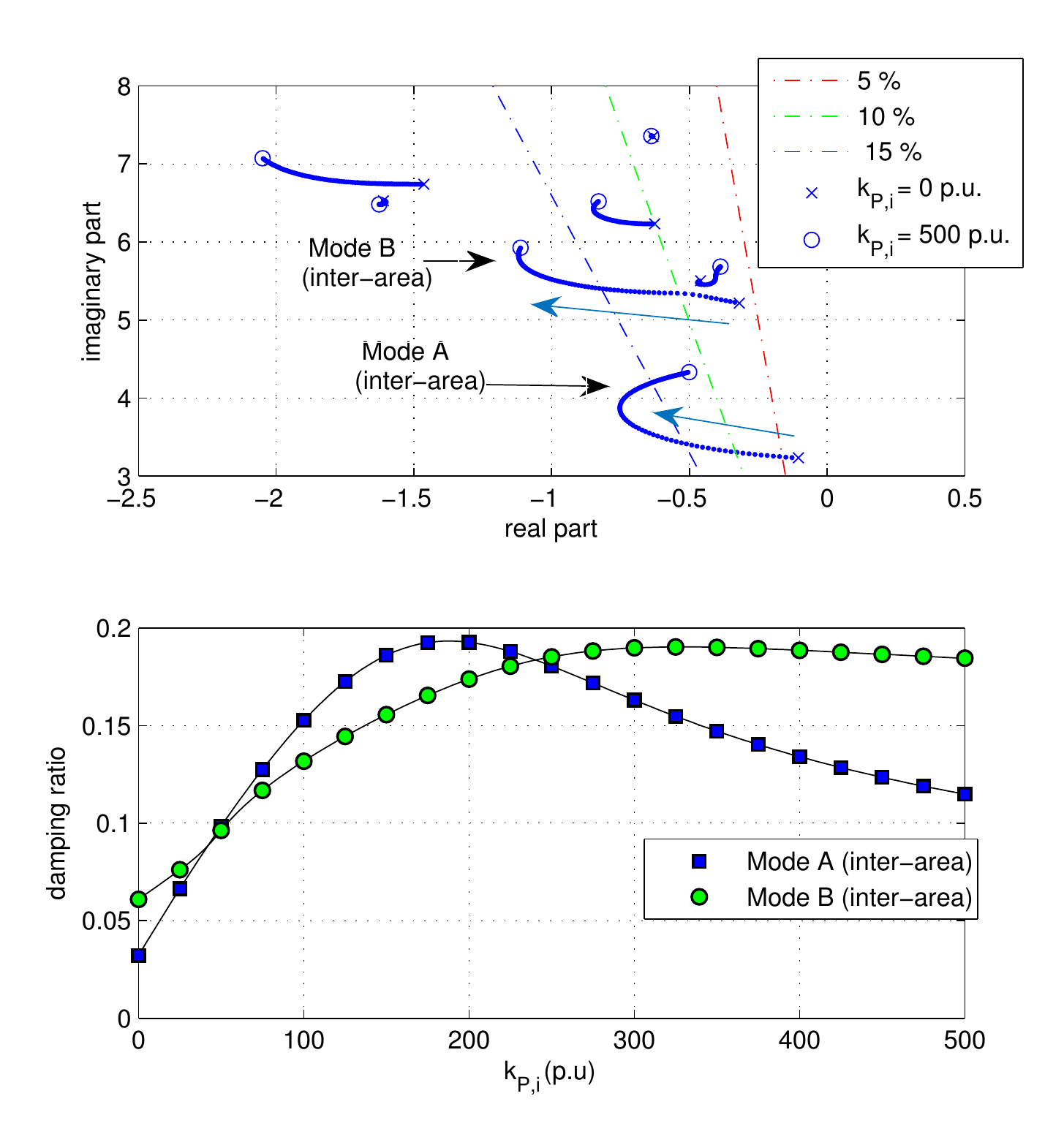}
\caption{Strategy P-WAF. Impact of gains $k_{P,i}$ on (upper plot) evolution of electromechanical modes and (lower plot) damping ratio.}
\label{fig:Nordic32A_ssa_pwaf_results}
\includegraphics[width=0.6\columnwidth]{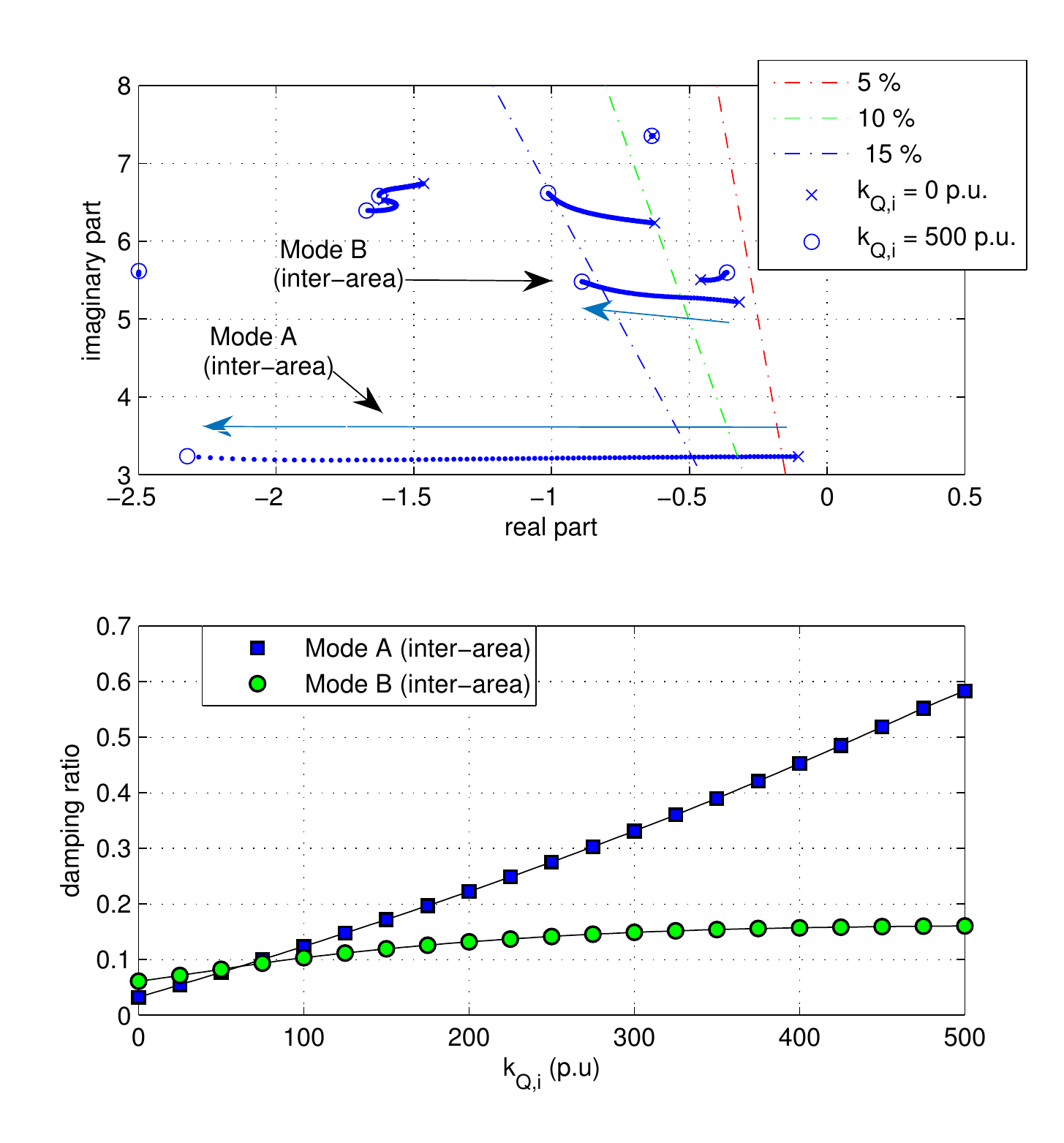}
\caption{Strategy Q-WAF. Impact of gains $k_{Q,i}$ on (upper plot) evolution of electromechanical modes and (lower plot) damping ratio}
\label{fig:Nordic32A_ssa_qwaf_results}
\end{center}
\end{figure}

Fig.~\ref{fig:Nordic32A_ssa_pwaf_results} shows the evolution of the inter-area modes of the system and their damping ratios as the gains of control strategy P-WAF (P injections) increase. Initially, inter-area mode A moves  towards  the left-hand side of the  complex plane as $k_{P,i}$ increases.  However, for high values of the controller gains this trend changes. The damping ratio of inter-area mode A increases as $k_{P,i}$ increases, it hits a peak for values around $k_{P,i}=200$~pu and it is reduced for higher values of the controller gains. Inter-area mode B moves  towards  the left-hand side of the  complex plane as $k_{P,i}$ increases and it starts to move up of the complex plane for high values of the controller gain. The damping ratio of inter-area mode B increases as $k_{P,i}$ increases and it approximately saturates around $k_{P,i}=200$~pu. The damping ratios of inter-area modes A and B are much higher than the ones obtained in the base case, for all values of the controller gain $k_{P,i}$. Notice also that no relevant negative impact of the controller on the damping ratio of other modes was observed.

Fig.~\ref{fig:Nordic32A_ssa_qwaf_results} shows the evolution of the inter-area modes of the system and their damping ratios as the gains of control strategy Q-WAF (Q injections) increase. Inter-area mode A moves  towards  the left-hand side of the  complex plane as $k_{Q,i}$ increases and its damping ratio increases, significantly. Inter-area mode B also moves  towards  the left-hand side of the  complex plane as $k_{Q,i}$ increases and its damping ratio also increases. Nevertheless, the damping ratio of mode  B is lower than the damping ratio of mode A. Furthermore, no relevant negative impact of the controller on the damping ratio of other modes was observed.

Results prove that with reasonable values of the controller gains for transient-stability improvement in strategies P-WAF and Q-WAF (e.g. $k_{P,i}=k_{Q,i}=200$~pu), inter-area modes are also damped successfully without affecting other modes significantly. Furthermore, small-signal stability techniques can be used to design the controller gains in order to obtain the required damping ratios of the electromechanical modes.

\FloatBarrier

\subsection{Non-linear time-domain simulation}\label{sec.results2_TDS}
\noindent The performance of  the control strategies has been tested by means of non-linear time domain simulation in PSS/E (electromechanical simulation), using the model proposed in~\cite{jrenedoPSSE2017}. 
Four cases are compared:
\begin{itemize}
	\item Base case: no supplementary control strategy.
	\item P-WAF: Strategy P-WAF (P injections) (Fig.~\ref{fig:P_supp_control}), with $k_{P,i}=200$~pu.
	\item Q-WAF: Strategy Q-WAF (Q injections) (Fig.~\ref{fig:Q_supp_control}), with $k_{Q,i}=200$~pu.
	\item PQ-WAF: Simultaneous modulation of P and Q injections with strategies P-WAF and Q-WAF, with $k_{P,i}=k_{Q,i}=200$~pu.
\end{itemize}
The rest of parameters of the control strategies are provided in~\ref{sec.control_strat_param}.

\begin{figure}[!htbp]
\begin{center}
\includegraphics[width=0.70\columnwidth]{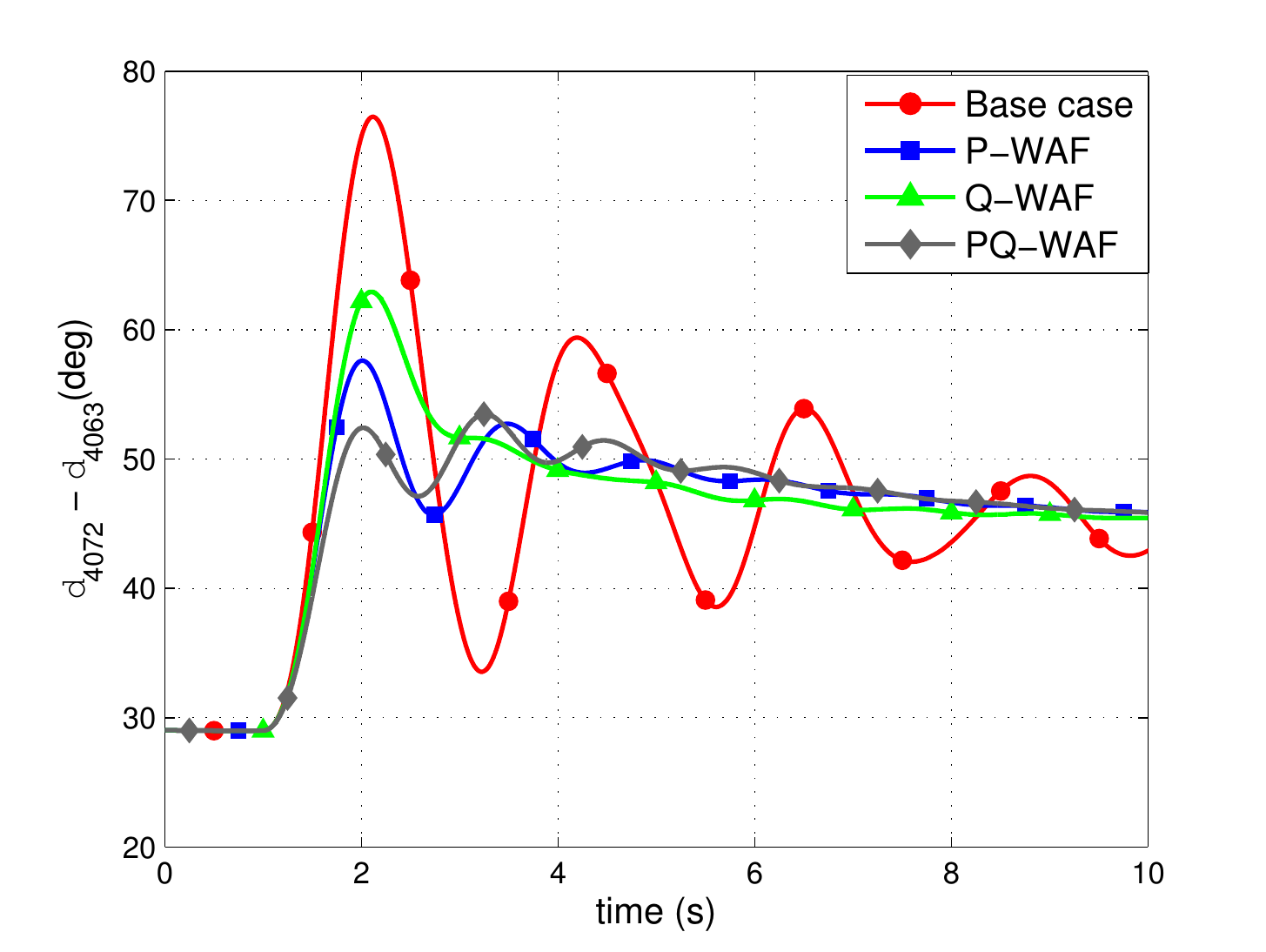}
\caption{Generator-angle difference.}
\label{fig:Nordic32A_sim1_Angles}
\end{center}
\end{figure}

Line 4012-4022 (see Fig.~\ref{fig:Nordic32A_mtdc}) is disconnected at $t=1$~s (a small disturbance). Fig.~\ref{fig:Nordic32A_sim1_Angles} shows the difference between the bus-voltage angles of generators 4072~(North) and 4063~(South). The three control strategies (P-WAF, Q-WAF and PQ-WAF) damp successfully the inter-area oscillations present in the base case. 

\begin{figure}[!htbp]
\begin{center}
\includegraphics[width=0.6\columnwidth]{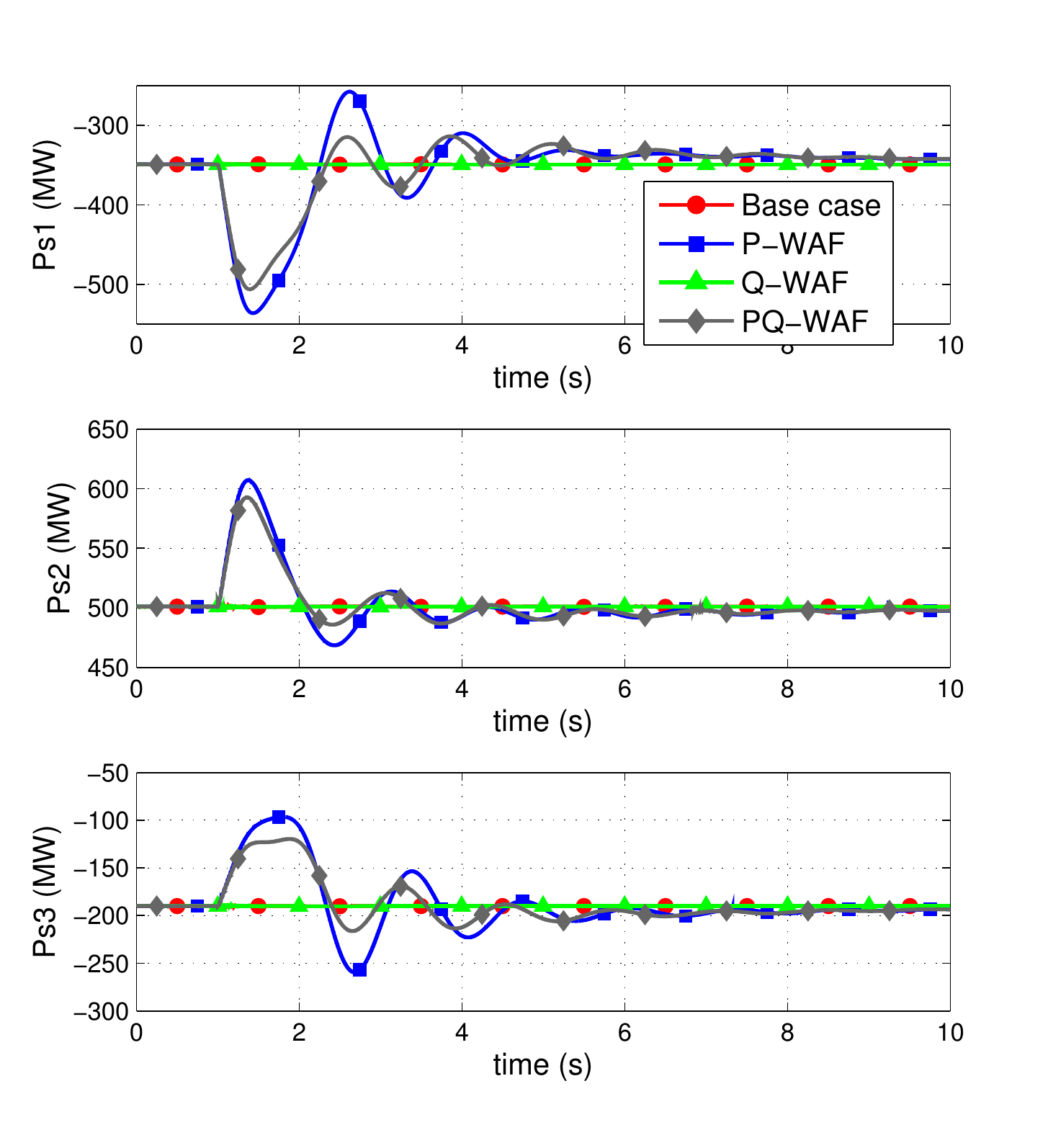}
\caption{Active-power injections of the VSCs ($P_{s,i}$).}
\label{fig:Nordic32A_sim1_Ps}

\includegraphics[width=0.6\columnwidth]{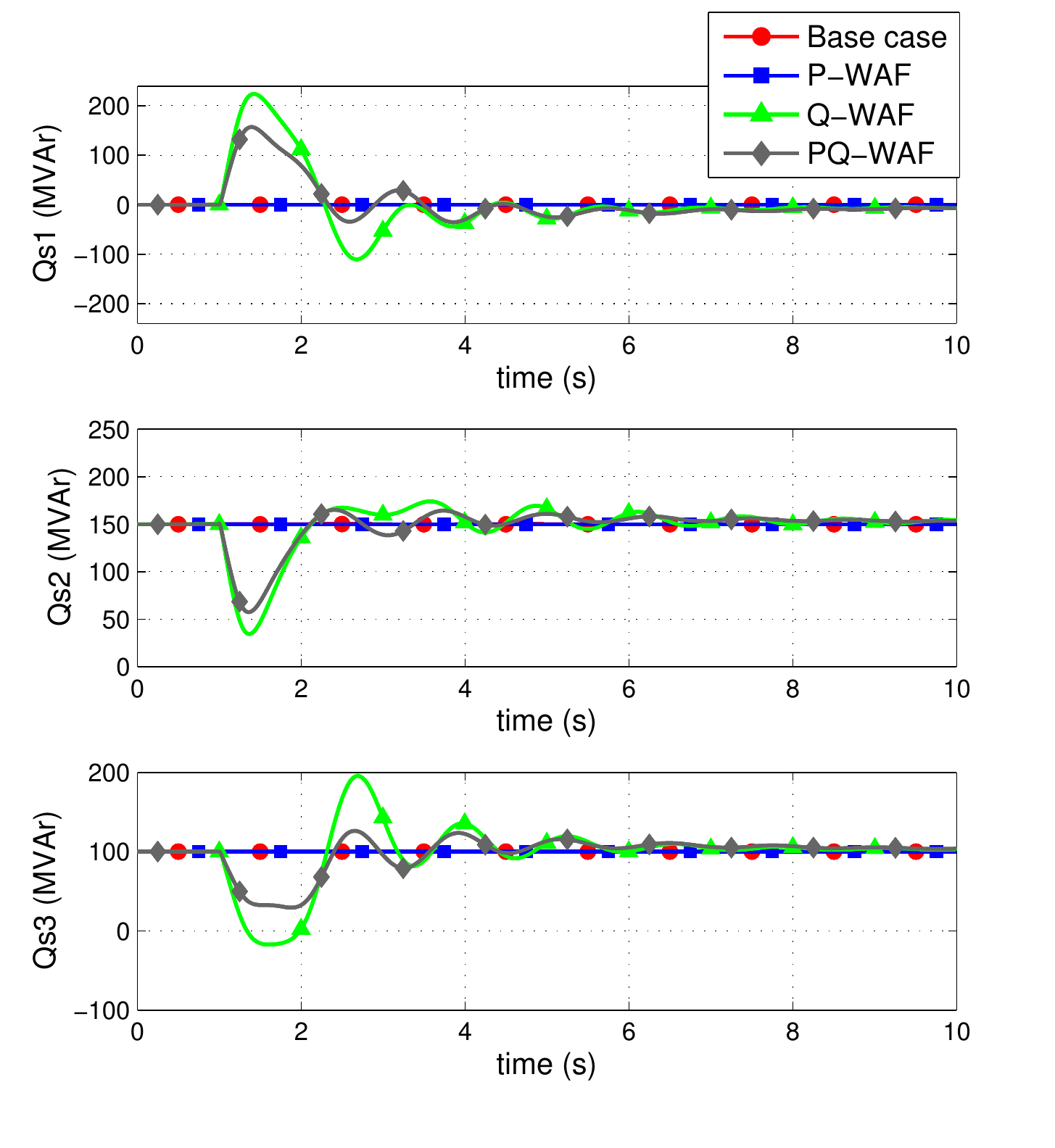}
\caption{Reactive-power injections of the VSCs ($Q_{s,i}$).}
\label{fig:Nordic32A_sim1_Qs}
\end{center}
\end{figure}

\begin{figure}[!htbp]
\begin{center}
\includegraphics[width=0.70\columnwidth]{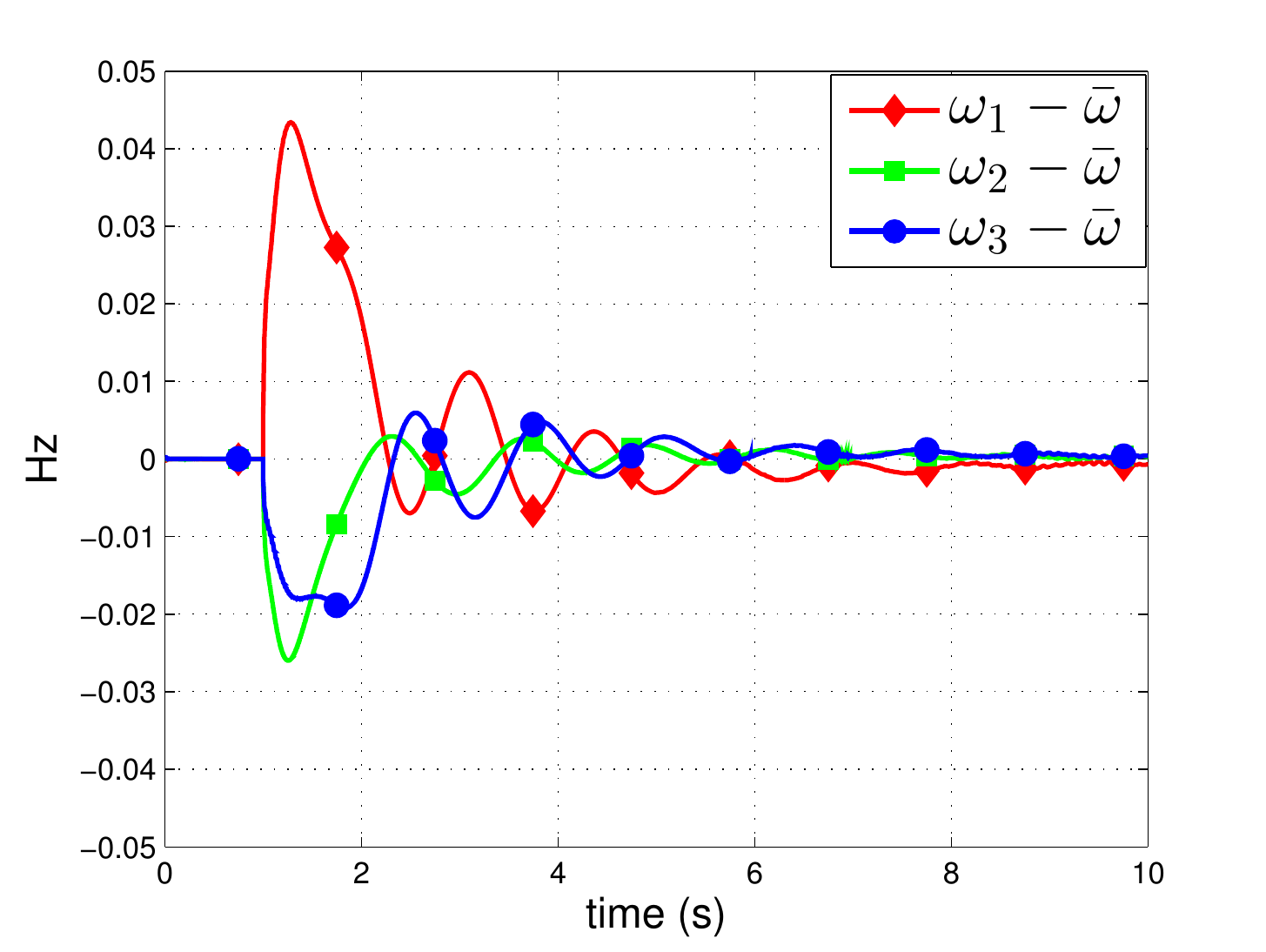}
\caption{Frequency deviations with respect to the WAF (case P-WAF).}
\label{fig:Nordic32A_sim1_freqs_dev}
\end{center}
\end{figure}

Figs.~\ref{fig:Nordic32A_sim1_Ps} and \ref{fig:Nordic32A_sim1_Qs} show the active-  and reactive-power injections of the VSC stations, respectively. In strategy P-WAF, only P injections are modulated. After the event, the P injection of VSC 1 increases because its frequency is above the WAF, while the P injections of VSCs 2 and 3 decrease because their frequencies are below the WAF (see Fig.~\ref{fig:Nordic32A_sim1_freqs_dev}, as an example). Analogously, in strategy Q-WAF, only Q injections are modulated. The behaviour of Q injections in Q-WAF is similar to the behaviour of P injections in P-WAF, but with opposite direction, due to the negative sign of Fig.~\ref{fig:Q_supp_control}. In strategy PQ-WAF, both, P and Q injections are modulated. The modulation of P and/or Q injections of the VSC stations contribute positively to damp inter-area oscillations. 

\FloatBarrier

\subsection{Overall performance}\label{sec.results2_comparison}

\noindent This section analyses the overall performance of the control strategies: their capability to improve (a) angle stability against large disturbances (transient stability) and (b) angle stability against small disturbances (electromechanical oscillations). Transient stability is the main application of control strategies P-WAF, Q-WAF and PQ-WAF and it has been analysed in detail in~\cite{javierr2016,iitcontrolQ2017}. Nevertheless, results on transient stability are also included as a reference in this paper to evaluate the overall performance of the control strategies. Transient-stability margin is quantified using the critical clearing time (CCT), which is defined as the maximum time duration that a fault can have without provoking loss of synchronism. An extremely severe fault is selected: a three-phase-to-ground short circuit applied to line 4031-4041a (close to bus 4041) (see Fig.~\ref{fig:Nordic32A_mtdc}), which is cleared by disconnecting the two circuits of the corridor (Fault I, for short). At the initial operating point, each circuit of the corridor carries 644.50~MW.

The impact of communication latency on the performance of the control strategies is also analysed by introducing a delay in the frequency set point calculated by each VSC ($\tau$) when calculating the WAF in (\ref{eq.weightedavinitial}): 
\begin{equation}\label{eq:freq_delay}
	\omega^{*} = \bar{\omega} e^{-\tau s}
\end{equation} 
The delay in~(\ref{eq:freq_delay}) has been implemented using a second-order Pad\'e's approximation, as proposed in~\cite{Yuwa2014}. Realistic values for the delays ($\tau= 50$~ms and $\tau= 100$~ms) were tested~\cite{Chow2015}.

Table~\ref{tab.results_comparison} shows damping ratios and frequencies of the inter-area modes A and B, and the CCT of Fault I, obtained in the base case and for control strategies P-WAF, Q-WAF and PQ-WAF. In all cases, controller gains are set to $k_{P,i}=k_{Q,i}=200$~pu and the rest of the parameters as described in~\ref{sec.control_strat_param}. The control strategies significantly increase the damping ratios of inter-area modes A and B, in comparison with the base case. The damping ratios slightly decrease in the presence of communication latency; however the improvements are still significant. The control strategies increase significantly the CCT of Fault I. Communication latencies do not reduce the CCT in strategy P-WAF. This is consistent with the results presented in~\cite{iit_coord_control_delay2019}: the DC-voltage droop attenuates the effect of communication latency when modulating P injections. However, the impact of communication latency is stronger when modulating Q injections with  strategy Q-WAF. Nevertheless, results obtained in the presence of communication latency are better than those obtained in the base case.

Hence, results prove that control strategies P-WAF, Q-WAF and PQ-WAF improve significantly both transient stability and electromechanical-oscillation damping.

\begin{table}[!htbp]
\begin{center}
\caption{Comparison. EO: Electromechanical oscillations (small disturbance), TS: Transient stability (large disturbance).}
\label{tab.results_comparison}
\scalebox{0.85}{
\begin{tabular}{|l|cc|c|}
\hline
 & \textbf{EO} &  & \textbf{TS}  \\ 
 \textbf{Case} & \textbf{Mode A} & \textbf{Mode B} & \textbf{Fault I}  \\
 & $\zeta$ (\%), $f$ (Hz) & $\zeta$ (\%), $f$ (Hz) & CCT (ms) \\ 

\hline
Base case	& 	 3.23 \%, 0.51 Hz &  6.10 \%, 0.83 Hz & 70 ms	   	 \\ 
\hline
P-WAF, delay 0 ms	 &  19.27 \%, 0.62 Hz & 17.38 \%, 0.89 Hz&	270 ms   	 \\ 
P-WAF, delay 50 ms	&  19.29 \%, 0.62 Hz &	17.39 \%, 0.89 Hz &  270 ms 	 \\ 
P-WAF, delay 100 ms	&  19.30 \%, 0.62 Hz &	17.39 \%, 0.89 Hz & 270 ms	   	 \\ 
\hline
Q-WAF, delay 0 ms	 &	 22.24 \%, 0.53 Hz & 13.18 \%, 0.86 Hz &  230 ms	 \\ 
Q-WAF, delay 50 ms	&  19.37 \%, 0.54 Hz & 13.06 \%, 0.86 Hz & 180 ms	   	 \\ 
Q-WAF, delay 100 ms	&  16.09 \%, 0.54 Hz &	12.73 \%, 0.86 Hz & 120 ms	   	 \\ 
\hline
PQ-WAF, delay 0 ms	 &  22.74 \%, 0.71 Hz & 21.23 \%, 0.91 Hz &	 320 ms	   	 \\ 
PQ-WAF, delay 50 ms	&  19.21 \%, 0.70 Hz & 21.06 \%, 0.91 Hz &	300 ms		   	 \\ 
PQ-WAF, delay 100 ms	&  16.75 \%, 0.69 Hz & 21.06 \%, 0.90 Hz &	280 ms		   	 \\ 
\hline
\end{tabular}
}
\end{center}
\end{table}

\FloatBarrier

\section{Conclusions}\label{sec.conclusions}
\noindent This paper analysed the impact of transient-stability-tailored supplementary controllers in VSC-MTDC systems on electromechanical-oscillation damping. In the control strategies analysed, each VSC of the MTDC system compares its own frequency with the weighted-average frequency (WAF) of the MTDC system and it modulates its P injection, Q injection or both simultaneusly (P-WAF, Q-WAF and PQ-WAF, respectively). 

The conclusions obtained in this paper can be summarised as follows:
\begin{itemize}
	\item Control strategies P-WAF, Q-WAF and PQ-WAF can be tuned to (a) improve transient stability and to (b) damp inter-area electromechanical oscillations too.
	\item The control strategies are robust against communication latencies.
\end{itemize}

\FloatBarrier

\appendix

\section{Appendix: Data}\label{sec.appendix}

\subsection{Data of the test system}\label{sec.app_test_system}
\noindent Data of the original Cigr\'e Nordic32A test system can be found in~\cite{CIGREB_Nordic32A_1995,Karlsson2013} and a comprehensive description of the system can be found in~\cite{Nordic32A2013}. Some modifications were made in order to stress the system and to reduce the damping ratio of inter-area oscillations. The modifications made are detailed in~\cite{iit_vsc_mtdc_pod2021}. Coverter and HVDC grid parameters are prvided in Table~\ref{tab:Nordic32A_sim1_system_parameters}.

\subsection{Parameters of the control strategies}\label{sec.control_strat_param}
\begin{itemize}
	\item Strategy P-WAF: $k_{P,i} = 200$ p.u, $T_{f,i} = 0.1$ s, $T_{W,i} = 10$ s, $\Delta p_{max,i} = 1.0$ p.u and $\alpha_k=1/3$. The gains are in nominal p.u. Different values of $k_{P,i}$ are analysed in Section~\ref{sec.results2_ssa}.
	\item Strategy Q-WAF: $k_{Q,i} = 200$ p.u, $T_{f,i} = 0.1$ s, $T_{W,i} = 10$ s, $\Delta q_{max,i} = 1.0$ p.u,  $V_{TH,i}=0.75$~pu and $\alpha_k=1/3$. The gains are in nominal p.u. Different values of $k_{Q,i}$ are analysed in Section~\ref{sec.results2_ssa}.
	\item Strategy PQ-WAF: Strategies P-WAF and Q-WAF are implemented simultaneously. The same parameters of strategies P-WAF and Q-WAF are used.
\end{itemize}

\begin{table}[H]
\caption{Converter \& HVDC grid data} 
\begin{center}
\scalebox{0.87}{
\begin{tabular}{|l|c|}
\hline
\textbf{Parameters} &   \\ 
VSC's rating are base values for pu& \\ 
\hline
Rating VSC, DC voltage, AC voltage & 1000 MVA, $\pm 320$ kV, $300$ kV   \\
Configuration & Symmetrical monopole \\
Max. active (reactive) power & $\pm 1000$ MW ($\pm 450$ MVAr) \\
Max. current & 1 pu ($d$-axis priority)  \\
Max. DC voltage & $\pm 10$ \% \\
Max. modulation index ($m_{i}^{max} = \sqrt{\frac{3}{2}} \cdot \frac{V_{dc,B}}{V_{ac,B}}$) &1.31 pu  \\ 
Current-controller time constant ($\tau$) & 5 ms  \\
Connection resistance ($r_s$)/reactance ($x_s$) & 0.002 pu / 0.17 pu \\
(reactor + 300/400 kV transformer) &  \\
P-control prop./int. gains:   ($K_{d,p1}$/$K_{d,i1}$) & 0/0 \space \space (i.e. $i_{d,i}^{ref}=p_{s,i}^{ref}/u_{s,i}$)\\
Q-control prop./int. gains: ($K_{q,p1}$/$K_{q,i1}$) & 0/0 \space \space (i.e. $i_{q,i}^{ref}=-q_{s,i}^{ref}/u_{s,i}$)  \\
DC-voltage droop constant ($k_{dc,i}$) & 0.1 pu  \\
VSCs' loss coefficients ($a$/$b$) in pu & 11.033/3.464 $\times 10^{-3}$ pu \\
VSCs' loss coefficients ($c_{rec}$/$c_{inv}$) in pu & 4.40/6.67 $\times 10^{-3}$ pu \\
DC-line series parameters ($R_{dc,ij}$/$L_{dc,ij}$) &2.05 $\Omega$/140.10 mH\\
DC-line shunt capacitance ($C_{cc,ij}$) & $1.79 \mu$F \\
Eq. VSC capacitance ($C_{VSC,i}$) & 193.21 $\mu$F  \\
Total eq. DC-bus capacitance ($C_{dc,i}$) & 195.00 $\mu$F  \\
\hline
\end{tabular}
}
\label{tab:Nordic32A_sim1_system_parameters}
\end{center}
\end{table}

\section*{Aknowledgement}\label{sec.aknowledgement}

\noindent Work supported  by the Spanish Government and  MCI/AEI/FEDER (EU) under Project  Ref. RTI2018-098865-B-C31 and by Madrid Regional Government under PROMINT-CM Project  Ref. S2018/EMT-4366.




\end{document}